\RequirePackage{fix-cm}
\documentclass[twocolumn]{svjour3}          
\smartqed  
\usepackage{graphicx}
\usepackage{booktabs}
\usepackage{hyperref}
\usepackage{xcolor}
\usepackage[sort, numbers]{natbib}
\setcitestyle{square}
\begin{document}

\title{A proposal and evaluation of new timbre visualisation methods for audio sample browsers}

\thanks{This work is partly funded by the Natural Sciences and Engineering Research Council of Canada (NSERC) and the\textit{Fonds Nature et Technologies} of Quebec (FRQNT).}



\author{Etienne Richan         \and
        Jean Rouat 
}


\institute{Etienne Richan \at
             Université de Sherbrooke \\
             NECOTIS, CIRMMT\\
             ORCID : 0000-0002-3378-5794\\
              \email{etienne.richan@usherbrooke.ca}           
           \and
           Jean Rouat \at
             Université de Sherbrooke \\
             NECOTIS, CIRMMT\\
            ORCID: 0000-0002-9306-426X\\
            \email{jean.rouat@usherbrooke.ca}
}

\date{Received: date / Accepted: date}

\maketitle
\begin{abstract}
Searching through vast libraries of sound samples can be a daunting and time-consuming task. Modern audio sample browsers use mappings between acoustic properties and visual attributes to visually differentiate displayed items. There are few studies focused on how well these mappings help users search for a specific sample. We propose new methods for generating textural labels and positioning samples based on perceptual representations of timbre. We perform a series of studies to evaluate the benefits of using shape, color or texture as labels in a known-item search task.
We describe the motivation and implementation of the study, and present an in-depth analysis of results. We find that shape significantly improves task performance, while color and texture have little effect. We also compare results between in-person and online participants and propose research directions for further studies.


\end{abstract}

\section{Introduction}
\label{intro}
Modern sample libraries can contain thousands of synthesized or recorded sound samples. A common approach when searching for a sample is to filter the contents of the library based on keywords or categories and then audition the resulting samples one by one.


Several media browsers have been developed to accelerate this process by placing samples produced by query results in a scatterplot visualisation, or a \textit{starfield display} \citep{ahlberg_visual_1995}.
Some implementations allow users to specify what metadata or audio descriptors to use as the axes of the display \citep{font_design_2010, brazil_audio_2003}, while others use dimensionality reduction (DR) methods to project a high-dimensional set of auditory features to a 2D space \citep{frisson_audiometro:_2014, heise_soundtorch:_2008, font_corbera_freesound_2017}. The latter approach provides less meaningful axes, but can produce an effective clustering of similar sounds. This feature based generation of sample coordinates is often augmented by a visual labeling method. These visual labels can help the user to recognize types of sounds or sounds that they have already auditioned. In existing sample browsers, colors are mapped to timbre features \citep{frisson_audiometro:_2014,brazil_audio_2003, font_freesound_2013, font_corbera_freesound_2017}, and shapes are used to either distinguish categorical variables (e.g. instrument type) \citep{brazil_audio_2003} or to visualize time-varying features \citep{frisson_audiometro:_2014, heise_aurally_2009}. 

\subsection{A novel sample browser with textural labels}
We developed a sound sample browser which lets users visually label sounds using textural images.  It uses a pretrained neural network model \citep{li_universal_2017} for artistic style transfer \citep{gatys_image_2016} to synthesize visual labels for all samples in the library based on a reference set of sound-image pairs. Users can simply choose textural images that they wish to associate with certain sounds and the software takes care of the rest. The advantage of using this method is that no explicit mapping between sound descriptors and visual parameters needs to be defined. We think this is an interesting alternative to the more common approach of associating specific audio features to shape or color parameters. Our browser also uses dimensionality reduction of timbre features to place samples in the interface.  

\subsection{Research questions}
While we wish to evaluate the design choices of our sample browser, to what degree \textit{any} of these types of visual labels improve sound search is still an open question. We designed our study to address the following questions:

\begin{itemize}
\item Is there a difference between using color, shape or texture as visual labels in an audio sample browser?
\item Does our timbre feature based placement method assist search, and if so, are these visual labels effective when the information provided by placement is removed?
\end{itemize}

This article presents the design of the study as well as methods for generating visual labels and placing musical samples based on their timbre features. We provide in-depth analysis of results from a group of 15 participants who were recruited to perform the study in a controlled environment. We find that the placement method and shape labels improve participant efficiency, but do not significantly improve task completion time. We also compare these results with a second group of 14 participants who performed the study remotely, but find few commonalities between the two groups.


%
\section{Related work}
The field of information visualisation provides a rich resource of theory and guidelines for visual label design. Borgo et al. \cite{borgo_glyph-based_2012} provide an extensive review of glyph visualisation, a general form of label designed to communicate values visually. Chen and Floridi \cite{chen_analysis_2013} propose a taxonomy for four types of visual channels: geometric channels (e.g. size, shape, orientation), optical channels (e.g. hue, saturation, brightness, texture, blur, motion), topological and relational channels (e.g. position, distance, connection, intersection) and semantic channels (e.g. numbers, text, symbols).

While visualisation theory principles can be applied to arbitrary sources of information, visualisations of sound can benefit from visual metaphors that appeal to intuitive associations we might make between acoustic and visual properties. 

\subsection{Cross-modal correspondences for timbre visualisation}
\label{sound_vis}
Studies of cross-modal correspondences provide useful insights for audio interface design as they highlight associations between vision and audition that a large part of the population might intuitively understand.
The \textit{kiki-bouba} experiment \citep {wolfgang_kohler_gestalt_1947} is an early study of such correspondences which found that across cultures and ages, most people associate the vocalized word \textit{``bouba"} with rounded shapes and \textit{``kiki"} with pointed shapes. Recently, an investigation of the cross-modal correspondence of timbre and shapes \cite{adeli_audiovisual_2014} came to a similar conclusion with regard to musical timbre.``Soft" timbres were associated with round shapes, while ``harsh", brighter timbres were associated with spiky shapes. The same work also highlighted a tendency to associate the soft timbres with blues and greens and harsh timbres with reds and yellows. 

Giannakis and Smith \citep{giannakis_comparative_2006} studied correspondences between acoustic descriptors and visual parameters, furthering work begun by Walker  \citep{walker_effects_1987} on associations between pitch, loudness and visual features such as size, position and lightness. With \textit{Sound Mosaics}, they studied associations between synthesized timbres and textural images containing repeated elements with varying parameters such as coarseness, distribution, and granularity. They found strong associations between granularity and spectral compactness as well as between coarseness and spectral brightness.

Grill and colleagues performed a study highlighting several high-level perceptual qualities of textural sounds 
\citep{grill_identification_2011} and proposed visualisations \citep{Grill2012VisualizationOP} as well as methods for extracting descriptors \citep{grill_constructing_2012} for each one. 
Two of their proposed perceptual metrics, height and tonality (measuring whether a sound is more tone-like or noise-like), are quite similar to the timbral descriptors for brightness and sprectral flatness. Both were visualised using color: height was mapped to a range of hue and brightness ranging from bright yellow (high) to dark red (low) and saturation was mapped to tonality.

Berthaut et al. \citep{berthaut_combining_2010} as well as Soraghan \cite{soraghan_animating_2014} studied potential correlations between acoustic properties and those of animated 3D objects. The former found a preference for associating the spectral centroid with color lightness and tonality with texture roughness. The latter found that participants preferred to associate geometrical resolution with attack time, spikiness with the spectral centroid and visual brightness with the ratio of even to odd harmonics. Both found that a common preference among participants was much less obvious when multiple mappings were in effect.

When developing a sample browser incorporating timbre visualisation, designers can either decide on implementing a fixed subset of these acoustic to visual mappings or provide options for users to modify the mappings themselves. This second option may increase the tool's versatility, but is dependant on users' knowledge and interpretation of acoustic and visual descriptors. This is what inspired us to develop our sample browser with a simple method for users to associate textures and timbres by selecting pairs of images and samples.


\subsection{Relevant work in audio browsers}
\label{browsers}

Dimensionality reduction (DR) of low- and high-level audio descriptors is a common practice in audio browser research. A concise overview of commonly used DR methods in audio browsers can be found in \citep{Roma2019AdaptiveMO} and \citep{Stober2013IncrementalVO}. \textit{Islands of Music} \cite{Pampalk2002ContentbasedOA} and \textit{MusicMiner} \cite{Mrchen2005DatabionicVO} popularized using self-organizing maps (SOM) to organize music libraries into topographic maps of musical genres based on a large number of extracted low-level and high-level features. As songs are generally associated with visual metadata such as album covers and pictures of the artists, these can be used to visually differentiate and help users recognize specific songs.

CataRT \cite{schwarz_real_time_2006}, a tool for concatenative sound synthesis and exploration, presents sound grains (very short sound samples) in a starfield display. Originally allowing users to choose audio descriptors to define each axis and sample colors, it was later augmented with a combination of DR methods to assist sound search in large collections \citep{Schwarz_search}. This tool seems to have been influential in the design of recent drum sample browsers and sequencers. \textit{The Infinite Drum Machine} \citep{kyle_mcdonald_infinite_nodate} demonstrated the creative possibilities of visualising t-SNE DR of drum samples in a web-based drum machine, while \textit{XO} \citep{xln_audio_xo_nodate} is an example of a professional tool based on similar principles. Both tools use color and placement to differentiate sample timbre and allow users to select regions of the sample space to associate with specific beats in a rhythm sequence. Sample color is used to visualise the third dimension of the reduced space in \cite{kyle_mcdonald_infinite_nodate}, while \textit{XO} uses sample color to distinguish predicted drum types (e.g. kick, snare, cymbal).

Stober and N\"urnberger developed \textit{MusicGalaxy} \citep{Stober2010MusicGalaxyAM}, a music browser proposing an innovative solution to the commonly occurring issue with dimensionality reduction that some similar elements can be projected to different regions of the reduced space. When focusing on a specific song, its nearest neighbours in the high-dimensional feature space are made obvious by increasing their size. In subsequent user studies, this method compared favorably to the more common ``pan and zoom" method of navigating large collections.


The following section describes the sample browsers that are closest to ours in design that also incorporated user studies in their development.

\subsection{Studies of audio sample browsers}
\label{other_studies}

Heise et al.\citep{heise_soundtorch:_2008} developed \textit{SoundTorch}, which uses a SOM to organize environmental sound samples in 2D space. Participants preferred their method to a list-based interface. They later added a visualisation of the temporal envelope as the contour of each element \citep{heise_aurally_2009}, but did not study the effects of this additional visual information on the effectiveness of the tool.
 
Frisson et al. \citep{frisson_audiometro:_2014} developed AudioMetro, which uses t-SNE DR of audio features and a proximity grid to place sound samples in a starfield display. They also use color and shape labels to differentiate samples. They map the timbre descriptor for brightness to the color lightness channel, and the temporal evolution of sharpness to the contour of the visual labels. Their study mainly evaluated the effect of different methods of spatial organisation of the sound samples, and they offer little analysis of the effect of the labels. 
They remark that simply using DR would often result in overlapping samples, which they solved by displacing samples to points on a regular grid. We encountered the same problem but implemented a different solution using simulated springs to push samples apart (section \ref{design:coords}), an approach also found in \citep{Schwarz_search}.

In their master’s thesis, Font \citep{font_design_2010} presented the results of queries to the Freesound database \citep{font_freesound_2013} using a starfield display. Their study found that participants were most successful at finding sounds when they could choose sound descriptors as axes. This was compared to placement obtained via PCA DR which they found to perform worse than random placement.

None of these studies looked in-depth into the effect of their choices of visual labels, so we made measuring this effect the main objective of our study design.

\section{Study design}
\label{design}
The goal of our study is to determine whether and to what extent different types of labels (shape, color, and texture) help in the task of searching for a specific sample in a starfield display. The secondary objective is to evaluate the effectiveness of our dimensionality reduction based placement approach (section \ref{design:coords}). We designed a known-item search task that would allow us to test different combinations of labelling and placement methods. In order to obtain a baseline for comparison, we create baseline variants for labels and placement. The label baseline uses grey circles to represent each sample and the placement baseline assigns random coordinates to each sample.

\subsection{Task design and interaction}
\label{design:task}

\begin{figure}[ht]
  \centering
  \includegraphics[width=\linewidth]{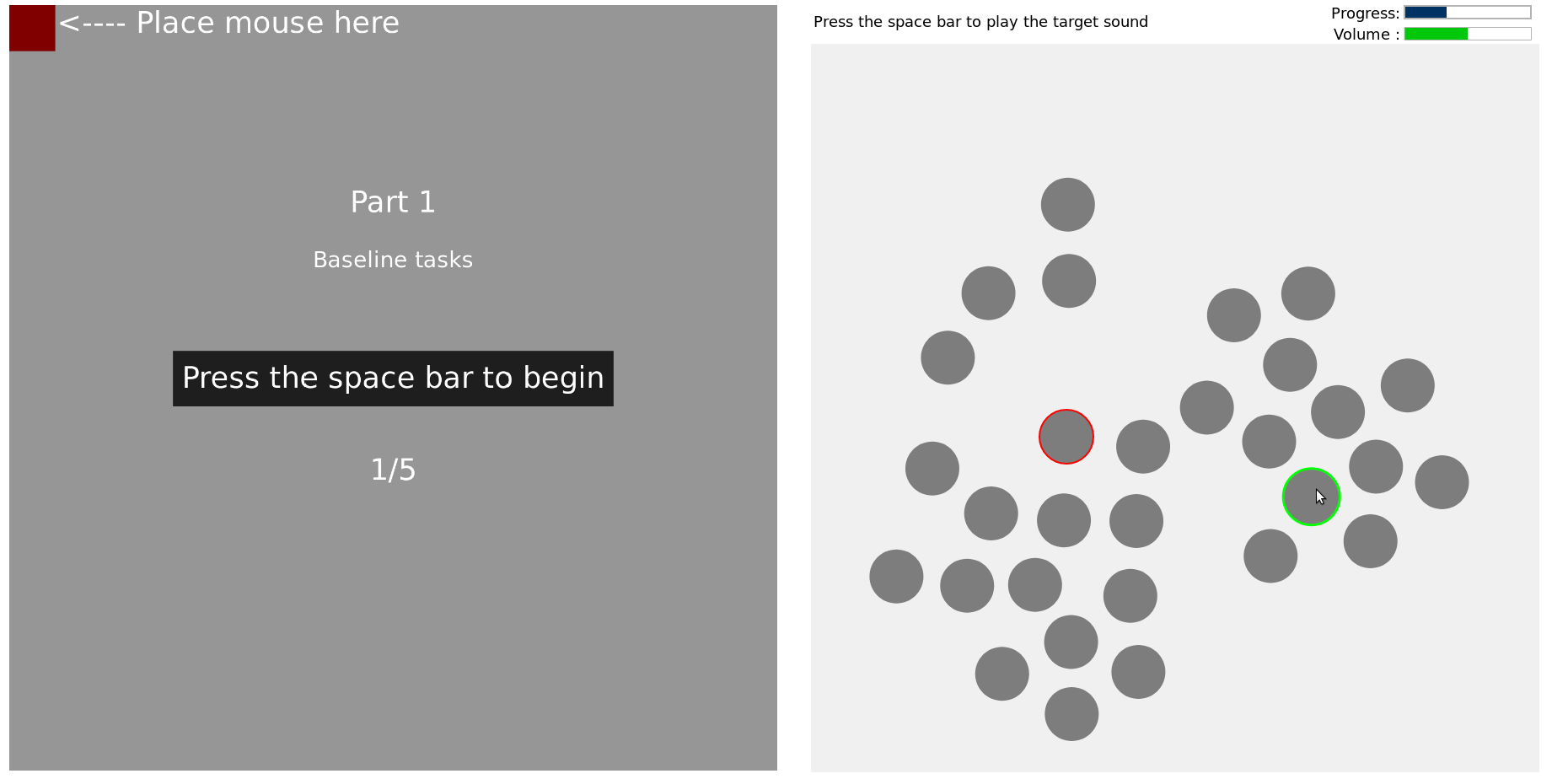}
  \caption{The task interface used in the study. \textit{Left} : The intermediary screen shown before each task. \textit{Right} : An example of a baseline task. Red outlines indicate an incorrectly clicked sound while a green outline highlights the current playing sound.}
\label{fig:interface}
\end{figure}

The task interface (with baseline labels) is shown in figure \ref{fig:interface}. In each task, 30 sounds are picked at random from a dataset and one is designated as the target sound the participant must find. Individual sound samples are displayed as circular shapes arranged on a light grey canvas. Samples are played by mousing over the corresponding element. Pressing the space bar plays the target sound and clicking on the correct element completes the task. Before each task, the participant must place their cursor in a corner of the canvas. This corner rotates clockwise around the canvas between tasks in order to vary the starting point. 

\subsection{Sets of tasks to introduce and evaluate timbre visualisation methods}
\label{design:task_sets}

Participants progress through the study by completing sets of tasks that introduce and evaluate the placement method and the visual labels. They first complete a practice task with baseline labels and random placement, that can be repeated until they are certain they understand how the interface works. They then complete a set of tasks with baseline labels and random placement. This represents the worst-case scenario, where no relevant information is being visualised. 

The rest of the study progresses by alternating between familiarization and evaluation tasks. During familiarization tasks, participants are encouraged to take their time and explore the set of samples while searching for the target sound. During evaluation tasks, participants are instructed to find the target sound as quickly as possible. We first introduce the dimensionality reduction (DR) based placement method with baseline labels. We then introduce a visual labelling method (color, shape or texture) with a set of tasks that uses the DR placement. Finally, we test the effectiveness of the labels on their own in a final set of tasks with the same labelling method and random placement. Before each set of tasks, participants read some brief instructions, which can be found in the supplementary materials of the paper (Online Resource 1, section 6).

The two placement methods (random and DR), and four label types (baseline, color, shape and texture) combine to form 8 different testing conditions. 
Participants also complete three survey-style questionnaires during the study referred to as $Q_0$, $Q_1$ and $Q_2$. In $Q_0$, participants provide basic demographic information (e.g. age, listening conditions, years of musical experience). In $Q_1$ and $Q_2$, participants are asked to rate the extent to which they used different search strategies for finding the target sound. They are also asked to rate whether the positioning or labeling of the sounds helped them in their search and how difficult they found the task overall. See the supplementary materials (Online Resource 1, section 5) for the full list of questions.

We designed the study to introduce and evaluate each type of label individually. Table \ref{table:basic_progression} summarises how a participant would progress through the entire study. For each set of task conditions, the same task will be repeated 5-10 times (depending on the task) with different samples. We designed the study to take approximately 30 minutes to complete. 

\begin{table}[h]
  \caption{ Series of tasks that participants complete while progressing through the study.
  \newline $P=$ Practice, $B_x$ = Baseline labels. $L_x$ = Color, shape or texture labels, $x_{DR}$=  Dimensionality reduction placement, $x_{R}$= Random placement
  }
  \label{table:basic_progression}
  \begin{tabular}{l l l l l l l l l }
    \toprule
    \textbf{Order} &1&2 &3&4&5&6 &7 &8  \\
    \midrule
    \textbf{Task type} &$Q_0$ &$P$ &$B_R$&$B_{DR}$&$L_{DR}$&$ Q_1$&$L_{R}$&$Q_2$ \\
    \bottomrule
\end{tabular}

\end{table}

\subsection{NSynth dataset and timbre feature extraction for sample placement and label generation}
\label{design:dataset}

We use musical samples from the NSynth dataset \citep{engel_neural_2017}, which consists of over 300,000 four-second samples produced by virtual instruments in commercial sample libraries. We are interested in differentiating timbre, so we use a subset of $\sim$800 samples with the same pitch and velocity. We use samples at midi note 64, which corresponds to musical note E4. 

For each sample, we use a cochlear filterbank\footnote{Available for download from \url{https://github.com/NECOTIS/ERBlet-Cochlear-Filterbank}} \citep{adeli_flexible_2016} to extract three profiles related to the perception of timbre: a spectral envelope, a roughness envelope (which measures perceived auditory roughness over time) and a temporal amplitude envelope. These timbre features are used both to determine sample placement as well as generate visual labels.

\subsection{Sample placement through dimensionality reduction} 
\label{design:coords}
We obtain two-dimensional coordinates for the musical samples through dimensionality reduction (DR) of the extracted features. We first use PCA to reduce each profile to a shorter feature vector, then apply UMAP \citep{mcinnes2018umap-software} to the concatenated vectors to produce a 2D arrangement of the samples that represents the distances between their high-dimensional timbre features. Recent related work \citep{frisson_audiometro:_2014,font_corbera_freesound_2017} has used t-SNE \citep{van_der_maaten_visualizing_2008} for dimensionality reduction, however, we find that UMAP obtains comparable results in significantly less computation time.


During testing, samples would occasionally overlap in the interface. To remedy this, before displaying each task, we run a brief physical simulation by placing virtual springs between samples causing them to push away from each other if they are overlapping.

\begin{figure*}[h]
  \centering
  \includegraphics[width=\textwidth]{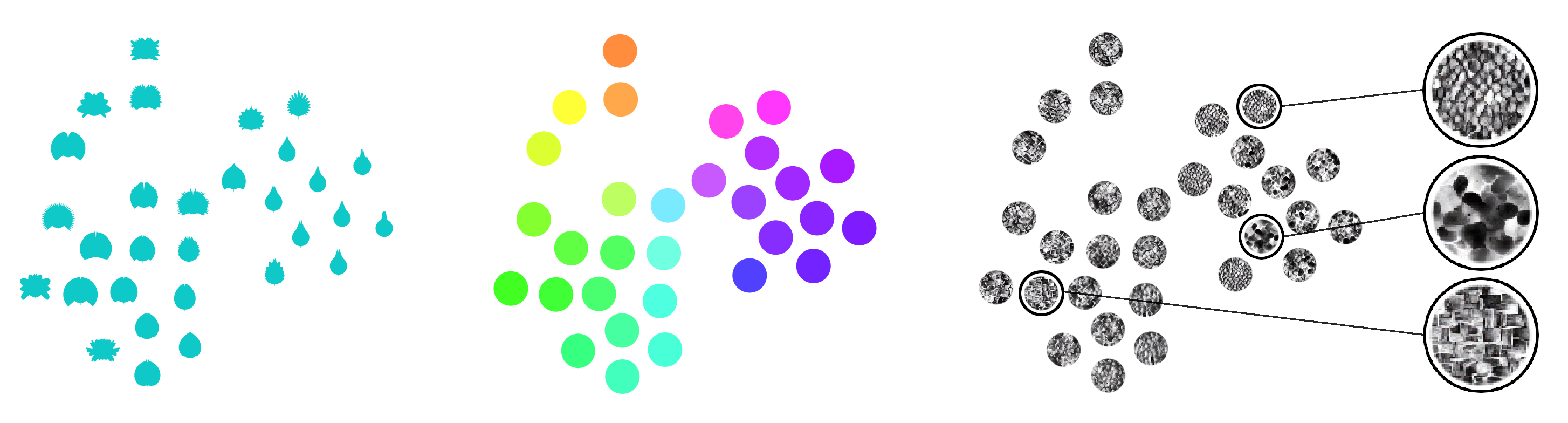}
  \caption{The three types of visual labels for the same set of randomly selected sound samples. From left to right: shape, color, texture.}
  \label{fig:methods}

\end{figure*}
\subsection{Visual label generation}
\label{design:labels}

Research on human visual perception has revealed that separate pathways are used to process shape, color and texture \citep{cant_independent_2008}. While this indicates that shape, color and texture could be used to distinguish between visual samples in a complementary fashion, this is not always the case. Different types of visual information can also interfere with each other \citep{callaghan_interference_1989}, so we chose to test each of these pathways separately. 

We are interested in differentiating timbre, which is often represented by continuous features describing some of the spectral or temporal characteristics of the sound. We generate labels that also vary continuously by using mappings from timbral features to visual parameters. Figure \ref{fig:methods} shows how the same set of samples would appear for each type of visual label. Labels could be of varying sizes, but in our studies, their diameter was 64 pixels.

\subsubsection{Shape}
\label{design:labels:shape}

We use the temporal envelope to generate our shapes because it visualises the attack time and low-frequency amplitude modulations in the signal, which are important timbral descriptors \citep{mcadams_perceptual_1995}.
We produce a unique shape for each sound by mapping the amplitude of the temporal envelope to the contour of a circular shape. We downsample the envelope to obtain a 20ms temporal resolution. This produces 200 distinct points for our 4 second samples. The radius of each point on a half-circle is described by equation \ref{eq:circle}.
\begin{equation}\label{eq:circle} 
R\left( {\theta } \right) = envelope[i],\,\,\,\theta  = \frac{i \cdot \pi }{{200}},\,\,\,i \in \left\{ {0,200} \right\}
\end{equation}

This half-circular shape is then rotated and mirrored along the x-axis to produce a symmetrical shape. The topmost point of the shape represents the amplitude of the envelope at the beginning of the sound and the bottom-most point the amplitude at the end of the sound.
This method is comparable to \citep{heise_aurally_2009}, with the main difference being that our approach produces symmetrical shapes, which are known to be easier to perceive and remember \citep{ward_multivariate_2008}.


\subsubsection{Color}
\label{design:labels:color}
For color, we use a simpler approach based on the coordinates of the samples in the reduced 2D space. The center of the spatial distribution of the entire set of samples is calculated and each sample is assigned a color based on its position relative to the center. The hue is determined by its angle relative to the center point and the saturation is determined by its distance to the center point. 
This can be imagined as laying a color wheel over the entire sample distribution and picking a color for each sample based on their location within it. Each sample's color reinforces the spatial information which is based on timbral similarity.

\subsubsection{Texture}
\label{design:labels:texture}
We developed a software tool to synthesize textural images for samples inspired by  Li et al.'s method for ``universal" style transfer  \citep{li_universal_2017}. The method is based on a pre-trained encoder-decoder neural network architecture (provided by \citep{li_universal_2017}). The encoder is an image classifier \citep{simonyan_very_2015}, while the decoder has been trained to reconstruct images from the activation patterns of the encoder. Noise is fed into the encoder network and the activation pattern is transformed to resemble that of a reference texture. The decoder can then produce a new image with textural properties that greatly resemble the original image. Our sample browser uses an optimized version of this architecture and provides a simple interface to extract, store, and interpolate between textural representations.

For the samples used in the study, eight medioids \citep{jin_k-medoids_2010} are found in the timbre space and each one is manually assigned a texture from the normalized Brodatz texture database \citep{abdelmounaime_new_2013}. We use black and white texture images in order to differentiate from the color method. We choose visually distinct textures for each medioid, and attempt to relate properties of the sounds to textural characteristics (e.g. a rapidly varying synth note is labeled with a chaotic rootlike texture, while a percussive mallet note is labeled with a texture of pebbles in an attempt to evoke the hardness of the material being struck). Textures for all the other sounds in the dataset are then produced by the texture synthesis method by interpolating between these textures based on their proximity to the medioids. 
Through this process, samples are assigned textural images whose visual properties vary in tandem with their timbral difference.


\subsection{Technologies used}
\label{design:technologies}
We use JATOS \citep{lange_just_2015} to build and host our study on a webserver, which allows us to easily distribute the study to participants using a web link.
JATOS studies are executed in the browser so no installation is required for the participants. 
The filterbank, dimensionality reduction and label generation are implemented in Python, and the resulting information is stored in a dictionary JSON file that is loaded by the web application. Pre-calculated shape and color information are stored as arrays in the dictionary. Each sample entry also points to a pre-generated JPEG texture file that is stored on the webserver. We use \textit{jsPsych} \citep{de_leeuw_jspsych:_2015} and \textit{p5.js} \citep{lauren_mccarthy_p5.js_nodate} to build the interactive components of the study and \textit{toxiclibs.js} \citep{kyle_phillips_toxiclibs.js_nodate} for the spring physics simulation. 

\subsection{Collected data }
\label{measures}

The measures we use to evaluate and compare labelling methods are summarized in table \ref{table:measure}. We collect several other data points from tasks, including the entire cursor trajectory, cursor speed, the number of misidentified samples and the number of times the participant listened to the target sound. The anonymised data collected in our studies will be available in the supplementary materials repository \citep{richan_timbre_2019}.
\begin{table}[ht]
  \caption{Recorded measures used to evaluate task performance}
  \label{table:measure}
  \begin{tabular}{l l l}
    \toprule
     \textbf{Measure}& \textbf{Description} &\textbf{Units} \\
    \midrule
    Time & Time taken to complete  &Seconds\\
    & the task&\\
    Hovered samples & Number of samples the & Count\\
    & cursor encountered & \\
    Distance & Total distance the mouse & Pixels\\
     & cursor travelled &\\
    \bottomrule

\end{tabular}
\end{table}

In summary, we designed our study to evaluate the effect of both visual labels and placement on searching for musical samples using technologies allowing for simple and easy distribution to participants. The studies we conducted using this design were approved by the \textit{Comité d’éthique de la recherche - Lettres et sciences humaines} of the University of Sherbrooke (ethical certificate number 2018-1795).

\section{Studies}
\label{results}

We conducted our study with 3 different groups of participants. The preliminary study did not lead to significant conclusions, prompting us to perform a study in a controlled environment with specifically qualified participants. Based on feedback from this second study, we decided to change the manner in which color labels were generated. We conducted a third iteration of the study with two objectives in mind : testing the new color labels and comparing the quality of results obtained in controlled and uncontrolled environments.

\subsection{Winter 2019: initial study}
\label{results:initial}
The precursor to this article \citep{richan_study_2019} summarized the results from a group of 28 computer engineering students. Students completed the study as part of coursework in a class on human perception and performed the study on their own computers. They were instructed to use a computer mouse and earphones. Based on those results, we concluded that it would be worthwhile to recruit a group of participants with a minimum of 2 years musical experience to perform the study in a controlled environment. We also realized the need for the baseline task with random placement, which was originally not part of the study. We only compared completion times in this study, and decided to record more data points in follow-up studies. 

\subsection{Summer 2019: qualified participants in a controlled environment}
\label{results:onsite}
15 participants were recruited from the music and engineering faculties of the University of Sherbrooke. They were required to have at least two years of recent experience working with sound in order to qualify them for the task of differentiating sounds by timbre alone. Participants completed the study in a secluded area on tablet-style laptops with a connected mouse and keyboard. The testing stations were equipped with Sennheiser HD 280 Pro headphones, connected via a Rega EAR amplifier and a Roland UA-1G USB interface.

Participants completed three passes of the study, with each iteration testing a different labelling method. Given that there are 3 types of labels, there are 6 permutations of the order in which they could be tested. These permutations were distributed between participants as evenly as possible.

\subsection{Summer 2019: new color labels and online participants}
A second group of participants, recruited in the same manner as the initial study, performed the study on their personal computers. For our analysis, we used results from 14 participants who reported more than two years musical experience and having completed the study in good listening conditions. 

We changed the way in which color labels were generated to better correspond other work in the field, particularly \cite{Grill2012VisualizationOP} and \citep{adeli_audiovisual_2014}. Our new method uses direct mapping from timbral descriptors to the hue and saturation of the circular labels. 
The spectral centroid (measuring timbral brightness) is mapped to a gradient from blue to red and spectral flatness (measuring tonality) is then mapped to the color's saturation. 
Shape and texture labels remain unchanged.

\section{Analysis methods}
\label{analysis_methods}

\subsection{Data transformation}
\label{results:transform}
Initial inspection of the collected measures (completion times, hovered samples and total mouse cursor distance) showed that distributions were quite heavily right-tailed. After performing Box-Cox transformations \citep{box_analysis_1964} on each set of measures, statistical models produce normally distributed residuals\footnote{For a linear regression model to be considered appropriate, the distribution of prediction errors (residuals) should resemble a normal distribution \citep{Martin17}.}.
Mean values and confidence intervals are calculated in transformed values and then back-transformed to their original units. Histograms of the collected measures and transformation parameters are included in the supplementary materials (Online Resource 1, section 2).

\subsection{Statistical models used to determine the effect of task conditions on performance}
\label{results:models}

We use general linear mixed-effect models as they support the repeated measures that characterize our study design and account for variance between participants. The placement method and label type are modeled as fixed effects and a unique identifier assigned to each participant is modeled as a random effect. We report estimated marginal (least-squares) means and confidence intervals. Analysis of variance (ANOVA) of fitted models estimates the probability of equal means.

For significance testing in survey responses, we use the Mann-Whitney U test \citep{mann_test_1947} when comparing between two groups and the Kruskal-Wallis one-way analysis of variance \citep{kruskal_use_1952} when comparing more than two groups.

\subsection{Packages and notebooks}
We provide R notebooks for reproducing our results in the supplementary materials repository \citep{richan_timbre_2019} and a list of R packages used in appendix \ref{appendix_r}.

\section{Results}
\label{results:intro}
We evaluate the effect of different labels and placement methods by comparing means of the recorded measures of task performance. We interpret P-values under 0.05 as strong evidence that the difference between means is significant (not due to chance) \citep{wasserstein_asa_2016}.

\subsection{Controlled study with qualified participants}
\label{results:b}
We first investigate the effect of the placement method, followed by that of the labels. Finally we present survey responses of interest.
\subsubsection{Effect of placement method}
\label{results:b:placement}

\begin{figure}[h]
  \centering
  \includegraphics[width=\linewidth]{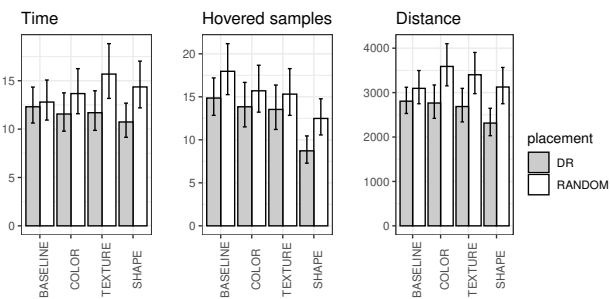}
  \caption{Means of task measures grouped by label type and placement. 95\% confidence intervals shown as error bars.}
\label{fig:group_means}
\end{figure}

Table \ref{table:mean_label_groups} shows the mean measures grouped by placement method and by label type. In figure \ref{fig:group_means} we plot these means with 95\% confidence intervals. 
We observe that for tasks with baseline labels, participants hovered 3 less samples with dimensionality reduction (DR) placement compared to random placement. Tasks with shape labels also show an $\sim$4 sample improvement with DR placement. For color and texture labels, while mean values of samples hovered are lower with DR placement, the difference is not statistically significant.
The differences in time and distance between placement methods are insignificant for basline tasks. For all label types aside from the baseline, the distance travelled by the mouse cursor with random placement is approximately 700 to 800 pixels longer than with DR placement. There is also a significant difference in completion times (approximately 3-4 seconds) in tasks with shape and texture labels.

\begin{table}[ht]
  \caption{Means of measures grouped by placement and labeling methods. Bold p-values indicate when the difference of mean measures between placement methods is significant.\newline DR = placement by dimensionality reduction.}
  \label{table:mean_label_groups}
  \begin{tabular}{l l l l l}
    \toprule
     \textbf{Measure}& \textbf{Label}& \multicolumn{2}{c}{\textbf{Placement}} & \textbf{p-value}\\
     \cmidrule(r){3-4}
      &&\textbf{DR} &\textbf{Random} & \\
    \midrule
    Time (s) &&&&\\
    & Baseline & 12.3 & 12.7 & 0.54\\
    & Shape & 10.7 & 14.3 & \textbf{0.0008}\\
    & Color & 11.6 & 13.7 & 0.051\\
    & Texture & 11.6 & 15.7 &\textbf{0.0005}\\
    Hovered&&&&\\
    samples& Baseline & 14.9 & 17.9 & \textbf{0.04}\\
    & Shape & 8.7 & 12.5 & \textbf{0.004}\\
    & Color & 13.8 & 15.7 & 0.16\\
    & Texture & 13.5 & 15.3 &0.30\\

    Distance &&&&\\
    (pixels)& Baseline & 2807 & 3094 & 0.12\\
    & Shape & 2313 & 3123 & \textbf{0.0005}\\
    & Color & 2765 & 3588 & \textbf{0.002}\\
    & Texture & 2681 & 3403 & \textbf{0.009}\\
    \bottomrule
\end{tabular}
\end{table}

\subsubsection{Effect of labeling methods}
\label{results:b:label}
We analyse the effect of labelling methods by comparing them to the baseline tasks. Table \ref{table:mean_diff} summarizes the differences between means. Our main takeaway from these results is that with shape labels, participants need to investigate significantly fewer samples before finding the target sound. Compared to the baseline, participants visited $\sim$6 less samples in tasks with shape labels before finding the target sound.

Times are not significantly changed when adding visual labels, except in the case of texture labels with random placement, where participants took $\sim$3 seconds longer to find the target sound.

\begin{table}
  \caption{Mean difference of measures for tasks with labels compared to baseline, grouped by placement method. Positive values indicate an improvement. Bold p-values indicate that the difference from the baseline is significant.\newline
    $\overline{B}$ = mean of measures with baseline labels \newline
    $\overline{L}$ = mean of measures with shape, color or texture labels \newline DR = placement by dimensionality reduction.
}
  \label{table:mean_diff}
  \begin{tabular}{l l l l l l}
    \toprule
     \textbf{Measure}&\textbf{Label} & \multicolumn{4}{c}{\textbf{Placement}} \\
     \cmidrule(r){3-6}
      &&\multicolumn{2}{c}{\textbf{DR}}& \multicolumn{2}{c}{\textbf{Random}} \\
      && \textbf{$\overline{B} -\overline{L}$} &\textbf{\textit{p-val.}}&\textbf{$\overline{B} -\overline{L}$}  &\textbf{\textit{p-val.}}\\
    \midrule
    Time (s) &&&&\\
    & Shape   & 1.59 & 0.18 &-1.58 & 0.26\\
    & Color   & 0.76 & 1.0 &-0.9  & 1.0\\
    & Texture & 0.73 & 0.96 &-2.91 & \textbf{0.03} \\
    Hovered&&&&\\
    samples& Shape & 6.1 & \textbf{4e-07} &5.5 & \textbf{0.002}\\
    & Color & 1.1 & 0.89  & 2.3 & 0.52 \\
    & Texture & 1.3 & 0.33  & 2.7 & 0.13 \\
    Distance &&&&\\
    (pixels) & Shape & 494 & \textbf{0.04} & -30 & 1.0 \\
    & Color & 41 & 1.0  & -493 & 0.20 \\
    & Texture & 120 & 0.94  & -307 & 0.61 \\
    
    \bottomrule
\end{tabular}
\end{table}


    



\subsubsection{Effect of iteration}
\label{results:b:iteration}
Given that participants complete the study multiple times, we are curious whether they improve at the different tasks over time.
Overall, we do not see a significant effect of iteration on measures. However, in the first set baseline tasks with DR placement, there is a significant difference between completion times (p=0.0078) and mouse speed (p=0.00020) compared to subsequent iterations. We hypothesize that participants were still becoming accustomed to using the interface during this first task in the study. 
There are no significant differences between iterations for tasks with random placement. When inspecting the data visually across iterations, there is a noticeable downward trend for tasks with labels and DR-based placement times, but the differences do not pass significance testing. 

\subsubsection{Questionnaire results}
\label{results:b:questionnaire}

The responses to many questions did not show significant differences between label types. This section highlights the most interesting responses. A full list of the questions can be found in the supplementary materials (Online Resource 1, section 5).

\begin{figure}[h]
  \centering
  \includegraphics[width=\linewidth]{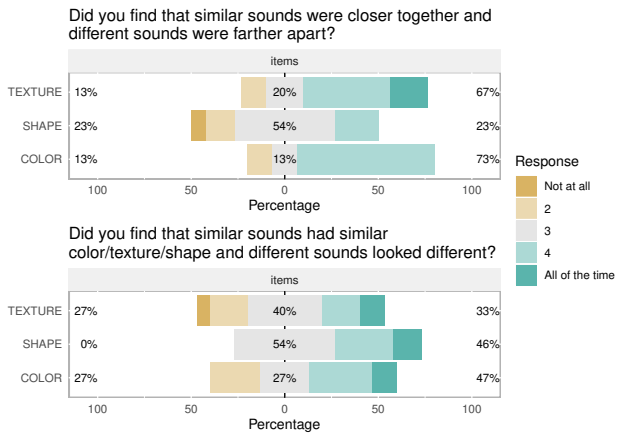}
  \caption{Responses to the questions about the perceived consistency of the labelling and placement methods }
  \label{fig:consistency}
\end{figure}

Figure \ref{fig:consistency} shows Likert plots of responses to questions about the perceived consistency of the placement of samples and the labelling methods in labelled tasks with DR placement. When rating label consistency, participants are quite evenly divided on texture labels, and lean towards color being more consistent. In the case of shapes, none rated their consistency below 3. When asked to rate whether similar sounds were located closer together, more participants responded with lower ratings after tasks with shape labels (p=0.032).

\begin{figure}[h]
  \centering
  \includegraphics[width=\linewidth]{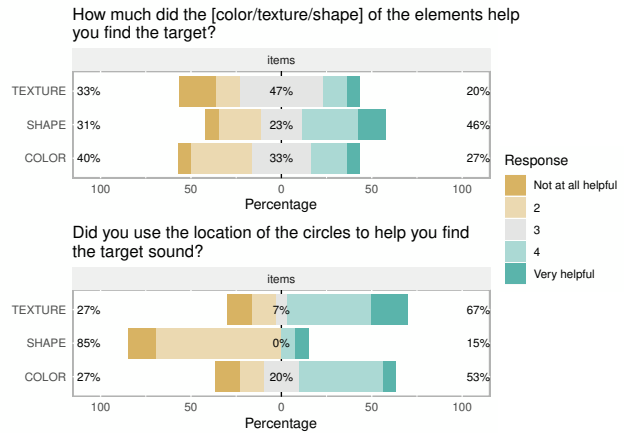}
  \caption{Responses to the questions about the perceived helpfulness of the labels and the sample placement}
  \label{fig:coordinates_help}
\end{figure}

Figure \ref{fig:coordinates_help} shows Likert plots of ratings of the perceived helpfulness of the placement of the samples during labeled tasks with DR placement. After tasks with color and texture labels, ratings skew towards the placement being helpful, but after tasks with shape labels they rate the placement as being less helpful (p= 0.04). This indicates that participants were paying less attention to the placement when shape labels were provided. 
Participants are quite evenly divided between positive and negative responses when rating the helpfulness of all of the label types.

\subsection{Online study}
\label{results:c}
Participants performed this study on their own computers, and were less experienced than participants in the previous study. While we used the same minimum experience criteria for both studies, 10 out of the 15 participants in the controlled environment study had at least 10 years of musical experience while 13 out of the 14 participants in this study had between 2 and 5 years of experience. 

\subsubsection{Effect of placement methods}
\label{results:c:place}

\begin{figure}[h]
  \centering
  \includegraphics[width=\linewidth]{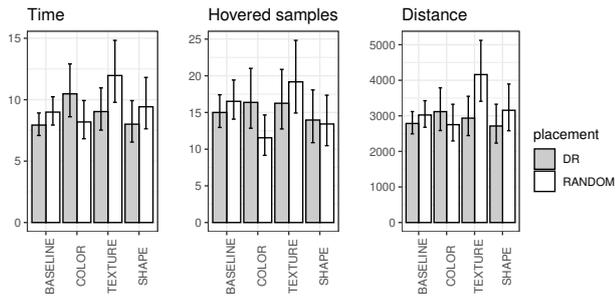}
  \caption{Means of measures grouped by label type and placement. 95\% confidence intervals shown as error bars.}
    \label{fig:group_means_class}
\end{figure}

Figure \ref{fig:group_means_class} shows mean measures grouped by placement and labeling method with 95\% confidence intervals. The only significant difference between placement methods is found in tasks with texture labels, where participants moved their cursor over $\sim$1200 more pixels searching for the target sound when positions were randomized. The exact means and p-values are provided in the supplementary materials (Online Resource 1, section 3, table S1).

    


\subsubsection{Effect of labeling methods}
\label{results:c:label}

Table \ref{table:mean_diff_online} shows the differences between labelled tasks and the baseline. There are two significant differences in completion times: on average, tasks with color labels and DR placement took 2.5 seconds longer than the baseline, while tasks with texture labels and random placement took 3 seconds longer than the baseline. 
5 less samples were visited on average in tasks with color labels and random placement compared to the baseline. Finally, when comparing mouse travel distance, in tasks with texture labels and random placement, participants covered $\sim$1100 more pixels than the baseline.

\begin{table}[h]
  \caption{Mean difference of measures compared to baseline, grouped by placement method. Positive values indicate an improvement. Bold p-values indicate that the difference is significant.\newline
      $\overline{B}$ = mean of measures with baseline labels \newline
    $\overline{L}$ = mean of measures with color, texture or shape labels
    \newline DR = placement by dimensionality reduction.
}
  \label{table:mean_diff_online}
  \begin{tabular}{l l l l l l}
    \toprule
     & & \multicolumn{4}{c}{\textbf{Placement}} \\
     \cmidrule(r){3-6}
      &&\multicolumn{2}{c}{\textbf{DR}}& \multicolumn{2}{c}{\textbf{Random}} \\
      \textbf{Measure}&\textbf{Label}& \textbf{$\overline{B} -\overline{L}$} &\textbf{\textit{p-val.}}&\textbf{$\overline{B} -\overline{L}$}  &\textbf{\textit{p-val.}}\\    \midrule
    Time &&&&\\
    & Shape   & -0.1 & 1.0 & -0.4 & 0.97 \\
    & Color   & -2.5 & \textbf{0.05} & 0.8  & 0.82 \\
    & Texture & -1.1 & 0.56 & -3.0 &  \textbf{0.05}\\
    
    Hovered&&&&\\  
    samples& Shape & 1.0 & 0.96 & 3.1 & 0.50 \\
    & Color & -1.4 & 0.92 & 5.0 & \textbf{0.05} \\
    & Texture & -1.3 & 0.93  & -2.7 & 0.71\\
    
    Distance &&&&\\
    & Shape & 71 & 0.99  & -131& 0.98 \\
    & Color & -333 & 0.72 & 272 & 0.82 \\
    & Texture & -150 & 0.95  & -1136 & \textbf{0.02} \\
    \bottomrule
\end{tabular}
\end{table}

    
    

\section{Discussion}
\label{discussion}
We designed our study to address the following two questions on the effectiveness of visual labels and sample placement in sound sample browsers.

\subsection{Does our timbre feature-based placement improve search, and if so, are visual labels still effective when the information provided by placement is removed?}

In the controlled study with qualified participants, we see significant differences in all three measures (completion time, hovered samples and cursor travel distance distance) when comparing our dimensionality reduction (DR) placement and random placement (table \ref{table:mean_label_groups}). In baseline tasks, the only significant effect of sample placement was a lowering of the number of hovered samples. However, in tasks with labels, we see a significant reduction in mouse travel distance when using the DR placement. This could be explained by participants jumping between visually similar labels when positions are random, in contrast to performing a sort of nearest-neighbour or grid-like search pattern with baseline labels. We expected completion times in baseline tasks to significantly differ between placement methods, but they did not. This indicates that the tasks with random placement were easier than we expected.

As to whether labels remained effective after placement information was removed, we saw that the number of hovered samples with shape labels was $\sim$6 samples lower when compared to the baseline (table \ref{table:mean_diff}). Additionally, the number of hovered samples is significantly lower when comparing DR and random placement within the shape label tasks ($\sim$4 less), so we can conclude that these two methods were complementary.

In the online study, we do not see the same effect of placement methods on participant performance. This could be explained by the fact that participants completed fewer tasks with DR placement overall and may not have had enough time to learn to use the placement effectively. This somewhat contradicts our conclusion from the analysis that the effect of iteration in the controlled environment study was negligible.

\subsection{Is there a difference between using color, shape or texture as visual labels
in an an audio sample browser?}
\label{discussion:label_effect}

In our controlled study with qualified participants, we found that shape labels significantly lowered the number of hovered samples compared to the baseline (table \ref{table:mean_diff}). We interpret the reduced number of hovered sounds as an improvement in participants' ability to visually differentiate samples, and thus avoid listening to irrelevant samples. Interestingly, this gain in efficiency did not translate to a significant gain in time, which could be explained by participants spending more time visually processing the scene. 
We did not find any differentiation between color labels and the baseline, so we can conclude that our approach to coloring samples in this study was ineffective. In their written comments, some participants expressed that the labels produced by the color mapping were in opposition with their preconceived associations between colors and timbres.
Textural labels did not differentiate from baseline tasks except in the case where participants took significantly longer to complete tasks with random placement. This indicates to us that the visual complexity of the textures was slowing them down. Many participants expressed that colored textures would have been much easier to differentiate.

The responses to the survey questions did not provide much additional insight into the differences between labeling approaches. They do however correlate with our previous observation that shape labels affected participants differently than the other two label types. 

In the online study, the lower number of hovered samples in tasks with colored labels and random placement (table \ref{table:mean_diff_online}) indicates that the new color labels could be helping participants find the target sound more effectively. This improvement is not present in the tasks with color labels and DR placement. In contrast with the previous study, shape does not seem to have much effect in reducing the number of hovered sounds when compared to the baseline.

\subsection{Comparing the controlled study to the online study}
\label{discussion:comparison}

Comparing both studies, the ranges of measures obtained are quite similar, but the measures from the online study have a higher variance than those from the controlled environment study (figures \ref{fig:group_means}, \ref{fig:group_means_class}), making it difficult to draw many conclusions from the results.

The difference in experience noted in section \ref{results:c} could explain why we do not find many commonalities in the results from the two studies and suggests that our minimum criteria for experience should be raised. Given that we do not observe similar effects of visualisation methods within the two groups, we are unsure whether we can recommend collecting data with this study in an uncontrolled environment. 
So far, our studies have been exploratory with small group sizes and our interpretations of results should be considered with this in mind. We remain optimistic that distributing this study online to a sufficient number qualified individuals would have a good chance of producing useful results.

\subsection{Further work }
\label{discussion:further_work}

In the studies we conducted, texture labels seemed to hinder participants more than help them. While we believe our new method for associating timbre to visual textures could prove useful for visualizing sounds in other contexts, future studies of this type of task could omit texture labels and concentrate on shapes and colors.
Further work could look into whether a combination of shape and color information can outperform shape alone. A significant advantage of using color is that it is much more tolerant to large changes in scale, while shapes need to be a minimum size in order to be visually distinct. 

Increasing the number of samples presented in each task would raise the overall difficulty and potentially accentuate the differences between the visualisation methods being tested.
We will likely also reduce the number of questions in future versions of the study, which would allow us to increase the number of tasks while maintaining the same overall duration. This study design could also be used to compare various methods of generating shapes. For example, we considered using the spectral envelope for shapes, as it contains other important timbral information such as the distribution of harmonic partials.
A variant of our study worth developing would allow participants to customize each visual labelling system. This would help mitigate some issues related to visual accessibility as our current colour schemes do not take colour blindness into account. It would also give participants the advantage of already understanding the underlying labelling system.

\section{Conclusion}
\label{conclusion}

We have conducted three studies using the study design presented in this article. Based on results from our first group, we decided upon minimal qualifications for participants and updated the study. Using the web based JATOS framework allowed us to quickly iterate on the study design and easily share it with our participants. Our second study brought qualified participants into a controlled environment and revealed a significant improvement when using shape labels, while texture and color labels did not provide noticeable advantages over the baseline. The final study produced few significant results, but provided some indications that our new colored labels are a step in the right direction.

Adding shape labels did not significantly improve task completion times, but did reduce the number of sounds visited before finding the correct one. It seems that the gain in efficiency of listening to less samples was offset by the extra time spent processing the additional visual information. We observed that this improvement persisted when information provided by the dimensionality reduction placement was removed, and that the two methods were complementary.

We consider our study design to have succeeded in allowing us to test a variety of placement and labelling methods and we were able to measure their individual and combined effects.
We hope other researchers will use our open-source implementation of the study\footnote{Available for download from \url{https://github.com/NECOTIS/timbre-visualisation-study}} as a starting point to pursue their own research questions related to sample browsers and sound search.

\section{Acknowledgements}
\label{acknowledgements}
This work is partly funded by the Natural Sciences and Engineering Research Council of Canada (NSERC) and the \textit{Fonds Nature et Technologies} of Quebec (FRQNT). We thank all of our participants for taking the time to complete our study.
We also thank our reviewers for their constructive feedback. Thanks to members of the NECOTIS laboratory of the University of Sherbrooke who beta-tested the study and provided feedback. We thank CIRMMT for providing access to their research infrastucture and travel funding. We also thank Frédéric Lavoie and and the GRPA of the University of Sherbrooke for generously lending us their testing facilities. Special thanks to Felix Camirand Lemyre for his advice on statistical modeling and analysis. 

%
%

\bibliographystyle{spbasic}      
\bibliography{bibliography.bib}   

\appendix{}
\section{R packages}
\label{appendix_r}
We use R \citep{r_citation} for our data analysis and figures. We use \textit{forecast} \citep{hyndman_forecast_manual} to estimate the optimal Box-Cox transform parameters as well as perform the forward and inverse transformations. We use general linear mixed-effect models from \textit{lme4} \citep{bates_lmer} and \textit{lmerTest} \citep{kuznetsova_lmertest}. Estimated marginal means and confidence intervals of fitted models are calculated with \textit{emmeans} \citep{emmeans_lenth}. Figures were produced with \textit{ggplot2} \citep{ggplot2} and \textit{likert} \citep{likert}. \textit{dplyr} \citep{dplyr} is used for data wrangling. 

%
%

\end{document}